\documentclass[11pt]{article}
\usepackage{amsmath,lscape,amsfonts,amssymb,amsthm,verbatim,color,lscape}
\usepackage[figuresright]{rotating}
\usepackage{url}
\usepackage{hyperref}
\hypersetup{colorlinks,
citecolor=black,
filecolor=black,
linkcolor=black,
urlcolor=black,
pdftex}

\usepackage{times}
\usepackage{blindtext}

\usepackage{sectsty}
\subsectionfont{\normalsize\bf}
\sectionfont{\large\bf}

\usepackage[font={footnotesize,it}]{caption}

\newcommand{\bea}{\begin{eqnarray}}
\newcommand{\eea}{\end{eqnarray}}
\def\bbf{\boldsymbol{\beta}}
\def\bhatbf{\widehat {\bbf}}

\def\bhatbf{\widehat {\bbf}}

\def\thbf{\boldsymbol{\theta}}

\def\rhbf{\boldsymbol{\rho}}
\def\Debf{\boldsymbol{\Delta}}
\def\E{{\rm E}\,}

\def\a{\mathbf{a}}
\def\g{\mathbf{g}}

\def\s{\mathbf{s}}

\def\V{\mathbf{V}}
\def\bfD{\mathbf{H}}

\def\R{\mathbf{R}}
\def\I{\mathbf{I}}
\def\M{\mathbf{J}}
\def\W{\mathbf{W}}
\def\x{\mathbf{x}}

\def\y{\mathbf{y}}
\def\Rtil{{\widetilde{\R}}}
\def\ahat{{\widehat\a}}

\def\O{{\boldsymbol\Omega}}

\def\D{{\Delta}}
\def\p{\partial}
\def\btil{{\widetilde{\bbf}}}

\begin{document}

\title{Correlation structure and variable selection in generalized estimating equations via composite likelihood information criteria}

\author{Aristidis~K.~Nikoloulopoulos\footnote{{\small\texttt{A.Nikoloulopoulos@uea.ac.uk}}, School of Computing Sciences, University of East Anglia,
Norwich Research Park, Norwich NR4 7TJ, UK}}
\date{}

\maketitle

\begin{abstract}

\noindent The method of generalized estimating equations (GEE) is  popular in the biostatistics literature for analyzing longitudinal binary and count data. It assumes  a generalized linear model (GLM) for the outcome variable, and a working correlation among repeated measurements.   In this paper, we introduce a viable competitor: the weighted scores method for GLM margins. We weight the univariate score equations using a working discretized multivariate normal model that is  
 a proper multivariate model. 
Since the weighted scores method is a parametric method based on  likelihood, we propose composite likelihood information criteria  as an intermediate step for model selection. The same criteria can be used for both correlation structure and variable selection. Simulations studies and the application example show that our method outperforms other  existing model selection methods in GEE. 
From the example, it can be seen that our methods 
not only improve on GEEs in terms of interpretability and efficiency, but also  can change the inferential conclusions with respect to GEE.
\medskip

\noindent \textbf{Key Words:}
AIC; BIC; Binary/Poisson regression; Composite likelihood; Generalized linear models; Weighted scores.
\end{abstract}

\section{Introduction}
\subsection{Motivating example}
A  European multi-center study has been conducted  to evaluate safety and efficacy for three fixed doses of a new drug in patients with major depressive disorder. 
Subjects were followed during 8 weeks  of treatment, starting from the beginning of the first week (baseline), and continuing with the beginning and the  end for the next seven weeks. Measurements were
taken at baseline (first week) and  every week during treatment resulting in a maximum of 8 measurements per subject (unequal cluster sizes).
The observations are coded as 1 if the Hamilton's depression (Ham-D) value is less than or equal to $80$ percent of the baseline value, and $0$ otherwise. The primary question of interest is  whether there is a  change in Ham-D rating from baseline to week 7, or the final visit in the case of those leaving the study early. The covariates  in this study are the treatment (active or placebo), and time in number of weeks from the baseline measurement. The study is described in full detail in \cite{sabo&chaganty10}.

\begin{table}[!h]
\centering
\caption{\label{hamd1}  
GEE estimates (Est.), along with their standard errors (SE) under different hypothesized correlation structures for the
Hamilton's depression data.}
\begin{tabular}{ccccccc}
\hline
Dependence           & \multicolumn{ 2}{c}{Independence} &                           \multicolumn{2}{c}{Exchangeable} &                           \multicolumn{2}{c}{AR(1)} \\\hline

Covariates           &           Est.  &         SE &      Est.  &         SE &      Est.  &         SE \\
\hline
     Intercept  &      -3.97 &       0.21 &      -3.67 &       0.24 &           -3.89 &       0.22  \\

      Treatment &      -0.54 &       0.21 &      -0.31 &       0.19 &          -0.53 &       0.20 \\

      Time &       0.97 &       0.05 &       0.95 &       0.06 &              0.96 &       0.06 \\
\hline
    $\rho$  &       0.00 &      -      &       0.24 &       0.22 &            0.48 &       0.22  \\
\hline
\end{tabular}
\end{table}

We decided to follow the generalized estimating equations (GEE) method    \cite{liang&zeger86,zeger&liang86} 
to obtain a population-averaged interpretation and to address the correlation between subject outcomes. 
Some practical correlation structures given the sparsity of the data (all  the responses are 0's at the baseline), are independence, exchangeable, and AR(1); for the unstructured case we noted divergence of the GEE estimates towards infinity, 
hence this structure is not considered in this example.
Table \ref{hamd1} gives the estimates  and standard errors  of the
model parameters obtained using 
GEE under different hypothesized correlation structures. The 
GEE estimates/robust standard errors are
calculated with  the R package 
{\it geepack} \cite{Halekoh&Hojsgaard&Yan2006}. 
It is obvious from the table that ignoring the actual correlation structure in the data could lead to invalid conclusions regarding the effect of treatment at reducing the depression levels in GEE analysis; e.g., for an exchangeable GEE analysis the treatment effect is statistically insignificant.

\subsection{Background}
The GEE method \cite{liang&zeger86,zeger&liang86}, which is  popular in biostatistics,
analyzes correlated data by assuming a generalized linear model (GLM)
for the outcome variable, and a structured correlation matrix to describe the pattern of association
among the repeated measurements on each subject or cluster. The correlations  are treated as nuisance parameters; interest focuses  on the statistical inference for the regression parameters. Recently,  Nikoloulopoulos and colleagues\cite{nikoloulopoulos&joe&chaganty10} developed the weighted scores method for regression models with dependent data.  The 
weighted scores method  is essentially an extension  of the
GEE approach, since it can also be applied to families that are not in
the GLM class. For concreteness,
the theory was illustrated for discrete negative binomial margins; that are not in the GLM
family.
Due to its generality, 
the theory also applies to GLM   margins.

In the case of dependent data with margins in the GLM family, the
weighted scores method is a competitor to GEE, but the weight matrices
are based on a plausible discretized multivariate normal (MVN) model, and the
parameters of the weight matrices are interpretable as latent correlation
parameters.  This avoids problems of interpretation in GEE for a working
correlation matrix that in general cannot be a correlation matrix of
the multivariate data as the univariate means change \cite{chaganty&joe06,sabo&chaganty10}.

The GEE is a proper methodology for regression with  dependent discrete margins in the GLM class if the variable selection in the mean function modelling
and the working correlation structure are correctly specified.
Hence, when conducting a GEE analysis, it is essential  to carefully model the correlation parameters, in order to avoid a substantive loss in efficiency in the estimation of the regression parameters \cite{Albert&McShane1995,Crowder95,wang&carey2003,Shults-etal2009}.
It turns out that model selection is  important  in longitudinal data analysis;  two practical  issues for modelling longitudinal data  are (a) the selection of the correlation (dependence) structure among various parametric correlation matrices, such as exchangeable,  AR(1), and unstructured, and (b) the variable/covariate selection in the regression model.

Two widely-used model-selection criteria are the Akaike's Information Criterion (AIC), and  the Bayesian  Information Criterion (BIC).  Since both are based on the likelihood and asymptotic properties of the maximum likelihood estimator, they cannot be used in GEE, which are based on moments with no defined likelihood.
There are some modifications of these criteria  in GEE;   they are not very powerful at choosing the correct correlation structure or the subset of covariates to be included in the regression model, possibly because they are not likelihood-based.
Pan \cite{pan2001} proposed the QIC criterion in GEE based on the quasi-likelihood constructed from the independent estimating equations.
Hin and Wang \cite{Hin&Wang2009} proposed a correlation information criterion (CIC),
which is just the penalty term of QIC, and showed that, without the term that is theoretically independent
of the correlation structures, CIC is more effective than QIC. Recently, Chen and Lazar \cite{Chen&Lazar2012}
applied empirical likelihood to the selection of working correlation structures in GEE, and obtained two correlation structure criteria, the empirical AIC (EAIC) and the empirical BIC (EBIC).  They have shown  that EAIC and EBIC are consistently  better than QIC and CIC.

The weighted scores method is a likelihood method and thus analogues of the AIC and the BIC for model and variable selection can be derived in the framework of the composite likelihood. 
Likelihood methods are  effective in selecting the best model from
a pool of candidates. This is a major advantage over GEE; GEE are based on moments and no likelihood
is defined, hence AIC and BIC cannot be derived. We propose/implement composite likelihood information
criteria, developed by \cite{Varin&Vidoni2005,Gao&Song2011}, as an intermediate step for correlation structure  and variable selection. The proposed criteria have  the similar attractive property with QIC of allowing covariate selection and working correlation structure selection using
the same model selection criteria, but at the same time being likelihood-based, we demonstrate that outperform all of  the aforementioned methods.

The remainder of the paper proceeds as follows. Section \ref{sec2} provides the theory of the weighted scores method for binary and Poisson regression with dependent data. Section \ref{cl1-sec} presents the composite likelihood  information criteria for model selection in the context of longitudinal data analysis with a GLM margin. Section \ref{sim-sec} describes the simulation studies we perform to  assess the performance of the composite likelihood information criteria in comparison with the existing criteria in GEE. In Section \ref{app-sec} we fully analyse the  Hamilton's depression data and show  a potential change of the inferential conclusions with respect to GEE.
We conclude with some remarks in Section \ref{disc-sec}, followed by a brief section with the software details and a technical Appendix.

\section{\label{sec2}Weighted scores method using GLM}
This section introduces  the theory of the weighted scores method for GLM regression with dependent data.
With GLM margins we have to deal only with univariate parameters that they are regression parameters, and thus we slightly differentiate ourselves from the general case in \cite{nikoloulopoulos&joe&chaganty10}. Before that, the first three subsections provide some background about important tools to form the weighted scores equations. These are the independent estimating equations in Subsection \ref{iee-sec}, the discretized multivariate normal (MVN) distribution in Subsection \ref{mvn-sec}, and the CL1 method \cite{zhao&joe05} in Subsection \ref{cl1ee-sec}.

\subsection{\label{iee-sec}Independent estimating equations}
For ease of exposition, let $d$ be the dimension of a ``cluster" or
``panel" and let $n$ be the number of clusters. The theory can
be extended to varying cluster sizes.
Let $p$ be the number of
covariates, that is, the dimension of a covariate vector $\x$ that
appears in the regression model is $p\times 1$.

Suppose that data are $(y_{ij},\x_{ij})$, $j=1,\ldots,d$, $i=1,\ldots,n$,
where $i$ is an index for individuals or clusters, $j$ is an index for
the repeated measurements or within cluster measurements.  The first component
of each $\x_{ij}$ is taken as ${1}$ for regression with an intercept.
The univariate marginal model for
$Y_{ij} $ is
$$
f_1(y_{ij}\; ;\; \nu_{ij})=\left\{\begin{array}{ccc}
h^{-1}(\nu_{ij})^{y_{ij}}\bigl\{1-h^{-1}(\nu_{ij})\bigr\}^{1-y_{ij}}& ,& Y_{ij}\sim \mbox{Bernoulli}\bigl\{h^{-1}(\nu_{ij})\bigr\}\\
 \frac{1}{y_{ij}!}\exp\bigl\{-h^{-1}(\nu_{ij})\bigr\}h^{-1}(\nu_{ij})^{y_{ij}}& ,& Y_{ij}\sim \mbox{Poisson}\bigl\{h^{-1}(\nu_{ij})\bigr\}
 \end{array},\right.
$$
where $h(\cdot)$ is the link function, i.e., $\nu_{ij}=\x_{ij}^\top\bbf=h(\mu_{ij})$ with $\mu_{ij}=E(Y_{ij})$.  The possible choices for the link function $h(\cdot)$ for binary (logit and probit) and Poisson regression
are given in Table \ref{derivatives}.

\begin{table}[!h]
\centering
\caption{\label{derivatives}The log-likelihood $\ell_1=\ell_1(\nu_{ij},y_{ij})$, its derivative
$\p \ell_1/\p \nu_{ij}$, and  the negative expectation of
the second derivative ${\p^2\ell_1}/{\p\nu_{ij}^2}$, i.e., $\D_{ij}^{(1)}$,  for Poisson, probit and logistic regression. Note that $\tilde\mu_{ij}=\mu_{ij}^{-1}(1-\mu_{ij})^{-1}$, $\tilde\phi=\phi\bigl(\Phi^{-1}(\mu_{ij})\bigr),$ where $\mu_{ij}=h^{-1}(\nu_{ij})$; $\phi$ and $\Phi$ denote the standard normal density and cdf, respectively.}
\begin{tabular}{ccccc}
\hline\\
Margin&$h$&$\ell_1$ &$\frac{\partial \ell_1}{\partial \nu_{ij}}$&$\D_{ij}^{(1)}$\\\\\hline\\
Poisson& $\log$ &  $-\log y_{ij}!-\mu_{ij}+y_{ij}\log \mu_{ij}$ & $y-\mu_{ij}$ & $\mu_{ij}$\\\\\hline\\
 & logit & & $y_{ij}-\mu_{ij}$ & $\mu_{ij}(1-\mu_{ij})$\\
 Bernoulli&&$y_{ij}\log\mu_{ij}+(1-y_{ij})\log(1-\mu_{ij})$ \\
&$\Phi^{-1}$&&$(y_{ij}-\mu_{ij})\tilde\mu_{ij}\tilde\phi$
&$\tilde\mu_{ij}\tilde\phi^2$\\
\hline
\end{tabular}
\end{table}

If for each $i$, $Y_{i1},\ldots,Y_{id}$ are independent, then the
log-likelihood is
\begin{equation}\label{L1}
L_1= \sum_{i=1}^n\sum_{j=1}^d\, \log f_1(y_{ij};\nu_{ij})=\sum_{i=1}^n\sum_{j=1}^d\,
 \ell_1(\nu_{ij},\, y_{ij}), 
\end{equation}
where $\ell_1(\cdot)=\log \, f_1(\cdot)$. The score equations for $\bbf$ are
\bea\label{gs3}
\frac{\p L_1}{\p \bbf}=
\sum_{i=1}^n\sum_{j=1}^d\, 
\frac{\p \nu_{ij}}{\p\bbf}  
\frac{\p \ell_1(\nu_{ij},\, y_{ij})}{\p\nu_{ij}}=\sum_{i=1}^n\sum_{j=1}^d\, 
\x_{ij}\s_{ij}^{(1)}(\bbf)  
=0.
\eea 
Let  $\x_{i}^\top=(\x_{i1}^\top, \ldots, \x_{id}^\top)$ 
and
$\s_i^{(1)\top}(\bbf)=
(\s_{i1}^{(1)\top}(\bbf), \ldots, \s_{id}^{(1)\top}(\bbf) )$, then 
the score equations (\ref{gs3}) can be written as
 \begin{equation}\label{independent score equations}
\g_1=\g_1(\bbf)=\frac{\p L_1}{\p \bbf}=  \sum_{i=1}^n\sum_{j=1}^d\x_{ij}^T\;\s_{ij}^{(1)}(\bbf)=
\sum_{i=1}^n\x_i^T\;\s_i^{(1)}(\bbf)=\mathbf{0}.
\end{equation}
The vector $\s_i^{(1)}(\bbf)$ has dimension  $d$,
since it comes from the derivatives with respect to $\nu$ for each member of a
cluster. The vector $\x_{ij}$ has dimension $p$ and the dimension of $\x_i$ is $d \times p$.

\subsection{\label{mvn-sec}Discretized MVN distribution}
The discretized MVN distribution (or MVN copula with discrete margins)
has been in use for a considerable length of time, e.g.  \cite{joe97}, and much earlier in the biostatistics \cite{Ashford&Sowden1970}, psychometrics \cite{Muthen1978}, and econometrics \cite{Hausman&Wise1978} literature.  It is usually known as a multivariate, or multinomial, probit model. The multivariate probit model is a simple example of the MVN copula with univariate probit regressions as the marginals. In the general case, the discretized MVN model has the following cumulative distribution function (cdf):
 $$F_d(y_1, \dots, y_d;\nu_1,\ldots,\nu_d,\R)=\Phi_d\left(\Phi^{-1}[F_1(y_1;\nu_1)],\ldots,
  \Phi^{-1}[F_1(y_d;\nu_d)];\R\right),$$
where $\Phi_d$ denotes the standard MVN distribution function with correlation
matrix $\R=(\rho_{jk}: 1\le j<k\le d)$, $\Phi$ is the cdf of the univariate standard normal,
and $F_1$'s are the univariate
cdfs of the marginal model for discrete data.

Implementation of the discretized MVN is feasible, but not easy, because the MVN distribution as a latent model for discrete response requires rectangle probabilities of the form 
\begin{equation}\label{MVNrectangle}
f_d(\y_{i})=\int_{\Phi^{-1}[F_{1}(y_{i1}-1;\nu_{i1})]}^{\Phi^{-1}[F_{1}(y_{i1};\nu_{i1})]}
\cdots \int_{\Phi^{-1}[F_{1}(y_{id}-1;\nu_{id})]}^{\Phi^{-1}[F_{1}(y_{id};\nu_{id})]}  \phi_d(z_1,\ldots,z_d;\R) dz_1\cdots dz_d
\end{equation}
where $\phi_d$ denotes the standard $d$-variate normal density  with correlation matrix $\R$.

\subsection{\label{cl1ee-sec}The CL1 method}
The MVN copula, although inherits the
dependence structure of the MVN distribution, lacks a closed form cdf;
hence likelihood inference might be difficult, as
$d$-dimensional integration is required for the computation of MVN rectangle probabilities \cite{Nikoloulopoulos&karlis07FNM,nikoloulopoulos13b,nikoloulopoulos2015}. When the joint probability is too difficult to compute, as in the case of the the discretized MVN model, composite likelihood is a good alternative \cite{varin08,Varin-etal2011}.

Zhao and Joe \cite{zhao&joe05} proposed the CL1
method to overcome the computational issues at the maximization routines for the MVN copula in a high-dimensional context.  Estimation of the model parameters
can be approached by solving the estimating equations obtained by the derivatives of the composite log-likelihoods.

In addition to the  the sum of univariate log-likelihoods (\ref{L1}) also consider the sum of bivariate log-likelihoods
$$L_2
\\=\sum_{i=1}^{n}\sum_{j<k}
\log{f_2(y_{ij},y_{ik};\nu_{ij},\nu_{ik},\rho_{jk})}=\sum_{i=1}^{n}\sum_{j<k}\ell_2(\nu_{ij},\nu_{ik},\rho_{jk};y_{ij},y_{ik}),$$
where
$$
f_2(y_{ij},y_{ik};\nu_{ij},\nu_{ik},\rho_{jk})=\int_{\Phi^{-1}[F_{1}(y_{ij}-1;\nu_{ij})]}^{\Phi^{-1}[F_{1}(y_{ij};\nu_{ij})]}
\int_{\Phi^{-1}[F_{1}(y_{ik}-1;\nu_{ik})]}^{\Phi^{-1}[F_{1}(y_{ik};\nu_{ik})]}  \phi_2(z_j,z_d;\rho_{jk}) dz_j dz_k;
$$ 
$\phi_2(\cdot;\rho)$ denotes the standard bivariate normal density  with correlation
 $\rho$. Note in passing the calculation of the bivariate normal rectangle probabilities $f_2(\cdot)$ is straightforward and does not involve additional computational effort as in the $d$-dimensional case.

Differentiating $L_1$ with respect to $\bbf$ leads to the univariate composite score function or independent estimating equations (\ref{independent score equations}). 
Differentiating $L_2$ with respect to $\R$ 
leads to the bivariate composite score function:
\begin{equation}\label{bivariate score equations}
\g_2=\frac{\p L_2}{\p \R}= \sum_{i=1}^{n}\s_i^{(2)}(\bbf,\R)= \sum_{i=1}^n\Bigl(\s_{i,jk}^{(2)}(\bbf,\rho_{jk}),1\leq j<k\leq d\Bigr)=\mathbf{0},
\end{equation}
where $\s_i^{(2)}(\bbf,\R)=\frac{\p\sum_{j<k}\ell_2(\nu_{ij},\nu_{ik},\rho_{jk};y_{ij},y_{ik})}{\p \R}$ and  $\s_{i,jk}^{(2)}(\bbf,\rho_{jk})=\frac{\p \ell_2(\nu_{ij},\nu_{ik},\rho_{jk};y_{ij},y_{ik})}{\p \rho_{jk}}$. The vector $\s_i^{(2)}(\R)$ has dimension $\binom{d}{2}$.

\subsection{Weighted scores assuming a working multivariate model}
The efficiency of estimating the regression parameters using the CL1 method can be low, since the method assumes independence. We will improve the efficiency by inserting weight matrices that depend on covariances of the scores assuming a ``working model", such as the discretized MVN \cite{nikoloulopoulos&joe&chaganty10}. 
The main idea is to 
weight the score equations in the case of independent data within clusters
or panels (\ref{independent score equations}), using a working model that is actually a proper multivariate model.

To this end, the weighted scores equations for Poisson/binary regression take the form:
\begin{equation}\label{wtee}
  \g_{1}^\star=\g_{1}^\star(\bbf)=\sum_{i=1}^n\x_i^T\,\W_{i,working}^{-1}\,\s_i^{(1)}(\bbf)=0,
\end{equation}
where   $\W_{i,working}^{-1}=
\Debf_i^{(1)}(\btil)\O_i^{(1)}(\btil,\Rtil)^{-1}$ with $\Debf_i^{(1)}(\btil)=\mbox{diag}(\Debf_{i1}^{(1)},\ldots,$ $\Debf_{id}^{(1)})$, $\Debf_{ij}^{(1)}=-\E(\frac{\p^2\ell_1}{\p\nu_{ij}^2})$ and $\O_i^{(1)}(\btil,\Rtil)=\mbox{Cov}(\s_i^{(1)})$ is the
the covariance matrix of $\s_i^{(1)}$ computed assuming a working discretized MVN model with estimation approached  via the CL1 method. 
The CL1 estimates $\btil$ for the regression parameters and $\Rtil$ for the correlation matrix  of the discretized MVN model  are obtained
by solving the CL1 estimating functions in (\ref{independent score equations}) and (\ref{bivariate score equations}), respectively.
 The matrices  $\Debf_i^{(1)}(\btil)$ and $\O_i^{(1)}(\btil,\Rtil)$ are $d\times d$ symmetric matrices.

The asymptotic covariance matrix of the
solution $\widehat{\boldsymbol{\beta}}$ in (\ref{wtee}) is
\begin{equation}\label{robust}
\V_{1}^\star=(-\bfD_{\g_{1}^\star})^{-1}\M_{\g_{1}^\star}(-\bfD^T_{\g_{1}^\star})^{-1},
\end{equation}
with
\bea\label{robust2}
  -\bfD_{\g_{1}^\star}
  =\sum_{i=1}^n\x_i^T\W_{i,working}^{-1}\Debf_i^{(1)}\x_i,\;
  M_{\g_{1}^\star} &=& \sum_{i=1}^n\x_i^T\W_{i,working}^{-1}
  \O_{i,true}^{(1)}(\W_{i,working}^{-1})^T\x_i,\nonumber
\eea
where $\W_{i,working}^{-1}=\Debf_i^{(1)}(\widehat{\boldsymbol{\beta}})\O_i^{(1)}(\widehat{\boldsymbol{\beta}},\Rtil)^{-1}$ and $\O_{i,true}^{(1)}$ is the true covariance matrix of $\s_i^{(1)}(\bbf)$. $\O_{i,true}^{(1)}$ can be estimated by the ``sandwich" covariance estimator, that is, $\s_i^{(1)}(\bhatbf)\s_i^{(1)^\top}(\bhatbf)$.
Explicit expressions
for the elements of the
log-likelihood $\ell_1=\ell_1(\nu_{ij},y_{ij})=\log f_1(\nu_{ij},y_{ij})$, its derivative
$\p \ell_1/\p \nu$, and  $\D_{ij}^{(1)}$, for  Poisson and binary regression are given in Table \ref{derivatives}.

To summarize, the steps  to obtain
parameter estimates and standard errors are the following:

\def\Rtil{{\widetilde{\R}}}

\def\atil{{\widetilde\bbf}}

\begin{enumerate}
\itemsep=0pt
\item Use a discretized MVN model
and estimate the parameters
using the CL1  method described in the previous subsection.
Let the CL1 estimates be $\atil$ for the univariate parameters
and $\Rtil$ for the correlation matrix.

\item  Compute the working weight matrices
$\W_{i,\rm working}=\O_{i}(\atil,\Rtil)\,\D_i(\atil)^{-1}$,
where $\O_{i}(\atil,\Rtil)$ is the
covariance matrix of
univariate scores based on the fitted discretized MVN model.

\item Obtain robust estimates $\bhatbf$ of the regression parameters
solving equations (\ref{wtee}) using ``working weight matrices"
$\W_{i,\rm working}$
using $\Rtil$, with a reliable non-linear system solver.

\item The robust standard errors for
$\ahat$ are  obtained calculating
the estimated covariance matrix $\widehat \V_1^\star$ by plugging
$\bhatbf$ for $\bbf$ and replacing
$\O_{i,\rm true}$ with $\s_i(\widehat\a)\s_i^T(\widehat\a)$.
This estimate is similar to the widely used ``sandwich" covariance estimator.
\end{enumerate}
The numerical methods used for various steps
have been implemented in the package {\tt weightedScores} \cite{nikoloulopoulos&joe11} within the open source statistical environment {\tt R} \cite{CRAN}; a wrapper {\tt R} function is also available.

\section{\label{cl1-sec}The CL1  information criteria}

As described in the preceding section, the CL1 method in \cite{zhao&joe05} is already used  to estimate conveniently  the univariate and latent correlation parameters of the discretized MVN model at the first step of the weighted scores method. Herein we also propose to use the CL1 method through its  information criteria, for correlation structure and variable selection in the weighted scores estimating equations. The remainder of the section proceeds as follows. 
Subsection \ref{godambe-sec} provides     the asymptotic covariance matrix for the estimator that solves the CL1 estimating equations. This is necessary   to form  the CL1 information  criteria in the  context of longitudinal data analysis with a GLM margin at Subsection \ref{criteria-sec}.

\subsection{\label{godambe-sec}Asymptotic covariance matrix}
The asymptotic covariance matrix for the estimator that solves (\ref{independent score equations}) and (\ref{bivariate score equations}), also known as the inverse Godambe information matrix \cite{godambe91}, is
\begin{equation}\label{inverseGodambe}
\V=(\bfD_\g)^{-1}\M_\g(\bfD_g^\top)^{-1},
\end{equation}
where $\g=(\g_1,\g_2)^\top$. First set $\thbf=(\bbf,\R)^\top$,  then

$$
-\bfD_\g=E\Bigl(\frac{\p \g}{\p\thbf}\Bigr)=
\begin{pmatrix}
E\Bigl(\frac{\p \g_1}{\p\bbf}\Bigr)&E\Bigl(\frac{\p \g_1}{\p\R}\Bigr)\\
E\Bigl(\frac{\p \g_2}{\p\bbf}\Bigr)&E\Bigl(\frac{\p \g_2}{\p\R}\Bigr)
\end{pmatrix}=
\begin{pmatrix}
-\bfD_{\g_1}&\mathbf{0}\\
-\bfD_{\g_{1,2}}&-\bfD_{\g_2}
\end{pmatrix},
$$
where $-\bfD_{\g_1}=\sum_i^n\x_i^\top\Debf_i^{(1)}\x_i$, 
$-\bfD_{\g_{1,2}}=\sum_i^n\Debf_i^{(1,2)}\x_i$, 
and
$-\bfD_{\g_2}=\sum_i^n\Debf_i^{(2,2)}$. 
Representations  of $\Debf_{i}^{(1)}$, $\Debf_{i}^{(1,2)}$, and, $\Debf_{i}^{(2)}$ for $d=4$ are given in an Appendix.

The covariance matrix $\M_\g$ of the composite score functions $\g$ is given as below
$$\M_\g=\mbox{Cov}(\g)=
\begin{pmatrix}\mbox{Cov}(\g_1) & \mbox{Cov}(\g_1,\g_2)\\
\mbox{Cov}(\g_2,\g_1) & \mbox{Cov}(\g_2)
\end{pmatrix}=
\begin{pmatrix}\M_\g^{(1)}& \M_\g^{(1,2)}\\
\M_\g^{(2,1)}& \M_\g^{(2)}\end{pmatrix}=
\sum_i\begin{pmatrix}\x_i^\top\O_i^{(1)}\x_i& \x_i^\top\O_i^{(1,2)}\\
\O_i^{(2,1)}\x_i& \O_i^{(2)}\end{pmatrix},
$$
where,
$$\begin{pmatrix}\O_i^{(1)}& \O_i^{(1,2)}\\
\O_i^{(2,1)}& \O_i^{(2)}\end{pmatrix}=
\begin{pmatrix}\mbox{Cov}\Bigl(\s_i^{(1)}(\bbf)\Bigr) & \mbox{Cov}\Bigl(\s_i^{(1)}(\bbf),\s_i^{(2)}(\bbf,\R)\Bigr)\\
\mbox{Cov}\Bigl(\s_i^{(2)}(\bbf,\R),\s_i^{(1)}(\bbf)\Bigr) & \mbox{Cov}\Bigl(\s_i^{(2)}(\bbf,\R)\Bigr)
\end{pmatrix}.$$
The dimensions of  $\O_i^{(1)}$, $\O_i^{(1,2)}$, $\O_i^{(2,1)}$, and, $\O_i^{(2)}$ are $d\times d$, $d\times \binom{d}{2}$, $\binom{d}{2}\times d$,  and, $\binom{d}{2}\times \binom{d}{2}$, respectively.

\subsection{\label{criteria-sec}CL1 AIC-BIC }
To this end, the CL1 AIC criterion  in \cite{Varin&Vidoni2005} and  the CL1 BIC criterion  in \cite{Gao&Song2011}  have the forms:
$$
\mbox{CL1AIC} =  -2L_2 + 2\mbox{tr}\Bigl(\M_\g\bfD_\g^{-1}\Bigr),
$$
$$
\mbox{CL1BIC} =-2L_2+\log(n)\mbox{tr}\Bigl(\M_\g\bfD_\g^{-1}\Bigr).
$$

To evaluate these criteria  at $\hat\thbf$ involves the computation  of matrices described above. The computations involve the trivariate and four-variate margins along with their derivatives.
In an Appendix we provide technical and computational  details for the calculation of $\M_\g, \bfD_g$ in (\ref{inverseGodambe}), 
and their  form for the case of one dependence parameter $\rho$.

\section{\label{sim-sec}Simulations}
In this section we 
assess the performance of the composite likelihood information criteria compared to the existing criteria in GEE.

We perform simulation studies to examine the reliability of using CL1AIC  and CL1BIC
to choose the correct model for longitudinal binary data. Similar results/conclusions can be expected for counts.
We compare the proposed criteria with the other available criteria in GEE.
In Subsection \ref{seccorsel} we assess the performance of CL1AIC, CL1BIC, QIC, CIC, EAIC, and  EBIC in correlation structure selection, and in  Subsection \ref{secvarsel} we investigate the performance of CL1AIC, CL1BIC, and QIC in variable selection.

\subsection{\label{seccorsel}Correlation structure selection}
We adopt the same model considered by  \cite{pan2001} and \cite{Chen&Lazar2012}.
We randomly generate $B=10^3$ samples of size {\bf $n = 50, 100, 200$} with $d=3$ using the R package {\it bindata} \cite{Leisch-etal-2012}  and logistic regression with
$p=3, \x_{ij}=(1,x_{1ij},j-1)^T$ where $x_{1ij}$ are taken as Bernoulli random
variables with probability of success $1/2$, and $\beta_0=0.25=-\beta_1=-\beta_2$.
In \cite{pan2001,Hin&Wang2009,Chen&Lazar2012} only structured matrices were considered  for the true correlation structures.
Here we consider both structured and unstructured matrices to allow for a comprehensive comparison.  We consider the following choices:

\begin{itemize}

\item  For exchangeable, we take
$\R$ as $(1-0.5)\I_3+ 0.5 \mathbf{O}_3$,
where $\I_3$ is the identity matrix of order $3$ and $\mathbf{O}_3$ is the
$3\times 3$ matrix of 1s.
\item For  AR(1), $\R$ is taken
as $(0.5^{|j-k|})_{1\le j,k\le 3}$.
\item For unstructured, we take $\R$ as
$$
\left( \begin{array}{rrr}
1.0 & -0.5 & -0.3 \\ -0.5 & 1.0 & 0.3 \\ -0.3 & 0.3 & 1.0
 \end{array} \right).
 $$

\end{itemize}
The latter  structure cannot be approximated by any of the aforementioned structured cases. This was not the case in \cite{Chen&Lazar2012}, who  used some stationary structures, which can be easily approximated by the exchangeable or AR(1) structure.

\setlength{\tabcolsep}{5pt}
\begin{table}[!h]
\centering
\caption{\label{corrsel}Frequencies of the correlation structure identified using the six different criteria, namely CL1AIC, CL1BIC, EAIC, EBIC, QIC, and CIC, from 1000 simulation runs in each setting. The first column indicates the true correlation structure  and its magnitude; IN, EX, and UN refer to independence, exchangeable, and unstructured correlation structure, respectively; the numbers of correct choices by each criterion are bold faced.}
    \begin{tabular}{lccccccccccccc}
    \hline
          &       & \multicolumn{4}{c}{n=50}      & \multicolumn{4}{c}{n=100}     & \multicolumn{4}{c}{n=200} \\
    \hline
          &       & IN    & EX    & AR    & UN    & IN    & EX    & AR    & UN    & IN    & EX    & AR    & UN \\\hline
    EX    & CL1AIC & 0     & \textbf{689} & 174   & 137   & 0     & \textbf{761} & 96    & 143   & 0     & \textbf{800} & 68    & 132 \\
    $\rho=0.5$ & CL1BIC & 0     & \textbf{825} & 146   & 29    & 0     & \textbf{891} & 92    & 17    & 0     & \textbf{923} & 67    & 10 \\
          & EAIC  & 0     & \textbf{692} & 131   & 177   & 0     & \textbf{792} & 53    & 155   & 0     & \textbf{848} & 2     & 150 \\
          & EBIC  & 0     & \textbf{794} & 138   & 68    & 0     & \textbf{917} & 59    & 24    & 0     & \textbf{987} & 7     & 6 \\
          & QIC   & 127   & \textbf{290} & 56    & 527   & 153   & \textbf{291} & 35    & 521   & 122   & \textbf{338} & 44    & 496 \\
          & CIC   & 0     & \textbf{183} & 54    & 763   & 0     & \textbf{197} & 22    & 781   & 0     & \textbf{203} & 2     & 795 \\\hline
    AR(1)    & CL1AIC & 0     & 92    & \textbf{749} & 159   & 0     & 31    & \textbf{767} & 202   & 0     & 3     & \textbf{703} & 294 \\
    $\rho=0.5$ & CL1BIC & 0     & 131   & \textbf{840} & 29    & 0     & 43    & \textbf{934} & 23    & 0     & 5     & \textbf{975} & 20 \\
          & EAIC  & 1     & 110   & \textbf{676} & 213   & 0     & 58    & \textbf{793} & 149   & 0     & 10    & \textbf{828} & 162 \\
          & EBIC  & 1     & 123   & \textbf{793} & 83    & 0     & 70    & \textbf{909} & 21    & 0     & 15    & \textbf{978} & 7 \\
          & QIC   & 100   & 161   & \textbf{241} & 498   & 105   & 183   & \textbf{270} & 442   & 86    & 142   & \textbf{295} & 477 \\
          & CIC   & 1     & 152   & \textbf{451} & 396   & 0     & 108   & \textbf{675} & 217   & 0     & 42    & \textbf{860} & 98 \\\hline
    UN    & CL1AIC & 0     & 22    & 2     & \textbf{976} & 0     & 0     & 0     & \textbf{1000} & 0     & 0     & 0     & \textbf{1000} \\
    $\rho_{12}=-0.5$ & CL1BIC & 0     & 120   & 6     & \textbf{874} & 0     & 8     & 0     & \textbf{992} & 0     & 0     & 0     & \textbf{1000} \\
    $\rho_{13}=-0.3$ & EAIC  & 8     & 17    & 0     & \textbf{975} & 0     & 0     & 0     & \textbf{1000} & 0     & 0     & 0     & \textbf{1000} \\
    $\rho_{23}=0.3$ & EBIC  & 69    & 59    & 4     & \textbf{868} & 2     & 4     & 0     & \textbf{994} & 0     & 0     & 0     & \textbf{1000} \\
          & QIC   & 41    & 121   & 20    & \textbf{818} & 34    & 120   & 20    & \textbf{826} & 30    & 121   & 17    & \textbf{832} \\
          & CIC$^1$   & 0     & 0     & 0     & \textbf{1000} & 0     & 0     & 0     & \textbf{1000} & 0     & 0     & 0     & \textbf{1000} \\
    \hline
    \end{tabular}
\begin{flushleft}
\begin{footnotesize}
$^1$: Since CIC does not account for penalty in terms of the number of correlation parameters estimated, its performance here is meaningless. 
\end{footnotesize}
\end{flushleft}
\end{table}

In Table \ref{corrsel}, we present the number of times that different
working correlation structures are chosen over 1000 simulation runs under each true correlation structure.
If the true correlation structure is exchangeable or AR(1), CL1BIC  improves remarkably over QIC, and is better that the other methods, especially for a small sample size, which is realistic for medical studies. The difference between the correct identification rate of CL1BIC and that of EBIC becomes small when the sample size increases to 100 or 200. If the true correlation structure is unstructured, CL1AIC and EAIC perform extremely well, behave similarly, and dominate the other methods.
For all correlation structures, the CL1AIC and EAIC criteria tend to choose the full-model correlation structure more often than CL1BIC and EBIC
do, since  AIC is more likely to result in an overparametrized
model than BIC in parametric settings \cite{Chen&Lazar2012}. 
Since CIC does not account for penalty in terms of the number of correlation parameters estimated, its performance is meaningless  when the number of correlation parameters is  different, i.e., when the true correlation structure is unstructured.

\subsection{\label{secvarsel}Variable selection}
We adopt the 
model considered by  \cite{pan2001}.
We randomly generate $B=10^3$ samples of size {\bf $n = 50, 100, 200$} with $d=3$ using the R package {\it bindata} \cite{Leisch-etal-2012}  and logistic regression with
$p=5, \x_{ij}=(1,x_{1ij},j-1,x_{3ij},x_{4ij})^\top$ where $x_{1ij},\beta_0,\beta_1,\beta_2$ are  as before, $x_{3ij},x_{4ij}$  are independent uniform random
variables in the interval $[-1,1]$ (and independent of $x_{1ij}$),   and $\beta_3=\beta_4=0$.
We consider the same candidate models with various  subsets of covariates, and include all the aforementioned parametric correlation structures, in addition to the exchangeable structure considered by \cite{pan2001}, as  true correlation structures. The subsets of covariates that we consider are the following:
\begin{itemize}
\itemsep=10pt
\item $\x_{1}=(1,x_{1ij})^\top$.
\item $\x_{12}=(1,x_{1ij},j-1)^\top$ (the true regression model).
\item $\x_{13}=(1,x_{1ij},x_{3ij})^\top$.
\item $\x_{123}=(1,x_{1ij},j-1,x_{3ij})^\top$.
\item $\x_{1234}=(1,x_{1ij},j-1,x_{3ij},x_{4ij})^\top$.
\end{itemize}

\setlength{\tabcolsep}{3pt}
\begin{table}[!h]
\centering
\caption{\label{varsel}Frequencies of the set of the variables  identified using the four different criteria, namely CL1AIC, CL1BIC, QIC, and QIC-IN, from 1000 simulation runs in each setting. The first column indicates the true correlation structure  and its magnitude; IN, EX,  and UN refer to independence, exchangeable, and unstructured correlation structure respectively;  the numbers of correct choices by each criterion are bold faced.}
\begin{small}
\begin{tabular}{lcccccccccccccccc}
\hline
           &            &                                     \multicolumn{ 5}{c}{Set $(n=50)$} &                                    \multicolumn{ 5}{c}{Set $(n=100)$} &                                    \multicolumn{ 5}{c}{Set $(n=200)$} \\\hline

&&$\x_1$&$\x_{12}$& $\x_{13}$& $\x_{123}$& $\x_{1234}$&$\x_1$&$\x_{12}$& $\x_{13}$& $\x_{123}$& $\x_{1234}$&$\x_1$&$\x_{12}$& $\x_{13}$& $\x_{123}$& $\x_{1234}$ \\
\hline
    EX    & CL1AIC & 103   & \textbf{517} & 26    & 183   & 171   & 23    & \textbf{626} & 13    & 179   & 159   & 2     & \textbf{673} &       & 181   & 144 \\
    
 $\rho=0.5$    & CL1BIC & 206   & \textbf{608} & 26    & 95    & 65    & 79    & \textbf{791} & 12    & 76    & 42    & 12    & \textbf{898} & 3     & 69    & 18 \\
          & QIC   & 247   & \textbf{449} & 71    & 119   & 114   & 105   & \textbf{624} & 32    & 122   & 117   & 11    & \textbf{726} & 5     & 157   & 101 \\
          & QIC-IND & 272   & \textbf{494} & 53    & 92    & 89    & 112   & \textbf{678} & 26    & 101   & 83    & 14    & \textbf{773} & 2     & 142   & 69 \\\hline
    AR(1) & CL1AIC & 166   & \textbf{452} & 54    & 164   & 164   & 65    & \textbf{562} & 25    & 173   & 175   & 10    & \textbf{613} & 5     & 204   & 168 \\
    $\rho=0.5$ & CL1BIC & 332   & \textbf{490} & 51    & 68    & 59    & 204   & \textbf{654} & 25    & 85    & 32    & 62    & \textbf{845} & 8     & 60    & 25 \\
          & QIC   & 383   & \textbf{361} & 87    & 79    & 90    & 215   & \textbf{523} & 47    & 109   & 106   & 63    & \textbf{691} & 14    & 140   & 92 \\
          & QIC-IND & 400   & \textbf{385} & 66    & 70    & 79    & 222   & \textbf{563} & 44    & 90    & 81    & 63    & \textbf{727} & 15    & 124   & 71 \\\hline
    UN    & CL1AIC & 284   & \textbf{328} & 100   & 122   & 166   & 204   & \textbf{423} & 60    & 150   & 163   & 76    & \textbf{557} & 20    & 170   & 177 \\
  $\rho_{12}=-0.5$   & CL1BIC & 540   & \textbf{289} & 70    & 46    & 55    & 464   & \textbf{399} & 44    & 58    & 35    & 273   & \textbf{628} & 19    & 56    & 24 \\
 $\rho_{13}=-0.3$    & QIC   & 447   & \textbf{281} & 120   & 68    & 84    & 349   & \textbf{401} & 73    & 87    & 90    & 170   & \textbf{606} & 33    & 116   & 75 \\
 $\rho_{23}=0.3$    & QIC-IND & 511   & \textbf{275} & 95    & 54    & 65    & 393   & \textbf{408} & 61    & 66    & 72    & 192   & \textbf{613} & 31    & 107   & 57 \\
    \hline
    \end{tabular}
    \end{small}
  \begin{flushleft}
\begin{footnotesize}
 $\x_{1}=(1,x_{1ij})^\top, \x_{12}=(1,x_{1ij},j-1)^\top \, (\mbox{the true regression model}),\, \x_{13}=(1,x_{1ij},x_{3ij})^\top, \x_{123}=(1,x_{1ij},j-1,x_{3ij})^\top,  \x_{1234}=(1,x_{1ij},j-1,x_{3ij},x_{4ij})^\top.$
  \end{footnotesize}
  \end{flushleft} 
\end{table}

In Table \ref{varsel}, we present the numbers of times that different
subsets of covariates are chosen over 1000 simulation runs under each true correlation structure. Among the criteria, we also use the QIC under independence (QIC-IN), which behaves better than QIC under the true correlation structure \cite{pan2001}.
If the true correlation structure is exchangeable or AR(1), CL1BIC performs better than QIC-IN, and its performance improves as the sample size increases. If the true correlation structure is unstructured, CL1AIC and QIC-IN behave similarly; the former behaves better for smaller sample sizes.
For all the candidate subsets of covariates,  CL1AIC  tends to choose the full-model more often than CL1BIC.

\section{\label{app-sec}The  Hamilton's depression data}

To select the appropriate correlation structure, we use the proposed model selection criteria in  the weighted scores estimating equations,  along  with the existing model selection criteria in GEE, based on the full model with all covariates and the interaction term between time and treatment (Table \ref{hamd2}, correlation structure selection). 

According to CL1AIC, CL1BIC, and QIC, the optimal correlation structure is the AR(1). 
CIC, EAIC, and EBIC 
are not able to reveal that and prefer the independence structure. Furthermore, with the CL1 criteria one can easily distinguish  between the various structures, since their difference in magnitude is large. This is not the case for the other criteria, where the differences are rather small. 
The AR(1) structure is plausible for this example, with measurements
that are approximately equally spaced in time, because it forces the correlation between
consecutive measurements on a subject to decrease with increasing separation in measurement
occasion.
For covariate selection, under the preferred AR(1) structure, we fit different models with different subsets of covariates, and find that the full model has the smallest CL1AIC, CL1BIC, and QIC-IND (Table \ref{hamd2}, variable selection).

\begin{table}[!h]
\centering
\caption{\label{hamd2}The values of the different criteria for model selection for the Hamilton's depression data. The smallest value of each criterion is boldfaced. }
\begin{tabular}{cccc}
\hline
\multicolumn{ 4}{l}{Correlation structure selection } \\\hline

  &         Independence &         Exchangeable &         AR(1) \\
\hline
    CL1AIC &   10194.88 (1.000)	&9832.90 (0.964) &	{\bf 9783.90 (0.960)}\\

    CL1BIC &   10212.42 (1.000)	&9866.94 (0.966)	&{\bf 9824.59 (0.962)}\\

      EAIC & {\bf 8568.37 (1.000)} &    8570.30 (1.000) &    8570.34 (1.000)\\

      EBIC & {\bf 8582.41 (1.000)} &    8587.84  (1.001) &    8587.89 (1.001)\\

       QIC & {\bf 1511.48 (1.000) } &    1537.87 (1.017) &    1513.56  (1.001)\\

       CIC &       8.46 (1.000) &       8.98 (1.061) & {\bf 8.41 (0.994)} \\
\hline
         \multicolumn{ 4}{l}{Variable selection } \\\hline

  &  Trt & Trt+Time &   Trt$\times$Time \\
\hline
CL1AIC-AR(1) &  17173.69 (1.000) &	9861.18 (0.574) &{\bf 	9783.90 (0.570)}\\

CL1BIC-AR(1) &   17200.81 (1.000) 	&9896.32 (0.575) &	{\bf 9824.59 (0.571)}\\

  QIC-IN &    2625.63 (1.000) &    1520.67 (0.579) & {\bf 1511.48 (0.576)} \\
\hline
\end{tabular}

\vspace{2ex}
\begin{footnotesize}
Relative terms  of the values of the different criteria  for each specific method under consideration are shown in parentheses.
\end{footnotesize}
\end{table}

Finally, Table \ref{hamd3} gives the estimates  and standard errors  of the
model parameters obtained using the  weighted scores estimating equations and GEE under the optimal AR(1) correlation structure. All estimates are similar and consistent. 
This example also shows that if the correlation structure and the variables  in the mean function modelling are  correctly specified, then there is no loss in efficiency in GEE.

\begin{table}[!h]
\centering
\caption{\label{hamd3} Weighted scores  and GEE estimates (Est.), along with their standard errors (SE) under the optimal correlation structure  and set  of covariates for the Hamilton's depression data.}
\begin{tabular}{ccccc}
\hline
Dependence            &                           \multicolumn{ 4}{c}{AR(1)} \\

Method           & \multicolumn{ 2}{c}{GEE} & \multicolumn{ 2}{c}{Weighted scores} \\
\hline
Covariates           &      Est.  &         SE &      Est.  &         SE \\
\hline
     Intercept  &      -4.67 &       0.32 &      -4.62 &       0.31 \\

      Treatment &       0.89 &       0.40 &       0.85 &       0.40 \\

      Time &       1.17 &       0.09 &       1.16 &       0.08 \\

  Treatment$\times$Time &      -0.36 &       0.11 &      -0.34 &       0.11 \\
\hline
   $\rho$ &       0.44 &       0.53 &       0.76 &      0.03      \\
\hline
\end{tabular}

\end{table}

\section{\label{disc-sec}Concluding remarks}
In this article, we have introduced binary and Poisson regression in the weighted scores method for regression with dependent data.  Our method of combining the univariate scores for binary and Poisson regression has the merit  of robustness to misspecification of the dependence structure like in generalized estimating equations, but with the additional advantage, with respect to generalized estimating equations, that dependence is expressed in terms of a ``real" multivariate model. We have compared our approach to GEE, and 
established our method as a viable competitor to GEE methods for both model selection and estimation.
The estimated correlation parameters in GEE cannot be interpreted, 
and sometimes violate the Fr\'echet bounds of the feasible range of the correlation \cite{chaganty&joe06}. This was the case in the application example as discussed by Sabo and Chaganty \cite{sabo&chaganty10}.
Our working MVN copula model is a proper multivariate model, and the correlations can be interpreted as latent correlations.

Comparing our method with  ML, one advantage  is that the weight matrices depend on covariances of the
scores; that is, only the bivariate marginal probabilities are needed. The ML method for the discretized MVN is feasible, but not easy, because multidimensional integration is needed to compute the MVN rectangle probabilities
\cite{nikoloulopoulos13b,nikoloulopoulos2015}.
Also  the weighted scores method is in a sense superior compared with the ML method;  based on a ``working" model 
 leads to unbiased estimating equations if univariate model is correct, while on the other hand ML estimates could be biased if the univariate model is correct but dependence is modelled incorrectly. This is the case in practice since the ``true''  model is not generally known. So weighted scores is robust to dependence if main interest is
in the univariate  parameters.

\section*{Software} 
R functions to implement the weighted scores method   and the CL1 information criteria for longitudinal binary and count data have been implemented in the package {\it weightedScores} \cite{nikoloulopoulos&joe11} within the open source statistical environment R \cite{CRAN}.

\section*{Acknowledgements}
We would like to thank Professor Harry Joe, University of British Columbia, for helpful comments and suggestions and Professor  Rao Chaganty, Old Dominion University,  for providing the Ham-D data.

\section*{Appendix}
In what follows we provide technical details for the calculation of the asymptotic covariance matrix for the estimator that solves the CL1 estimating equations, 
and its  form for the case of one dependence parameter $\rho$. This is the case for an exchangeable and AR(1) dependence structure, since  $\R$ is $(1-\rho)\I_d+ \rho \mathbf{O}_d$ ($\I_d$ is the identity matrix of order $d$ and $\mathbf{O}_d$ is the
$d\times d$ matrix of 1s) and $(\rho^{|j-k|})_{1\le j,k\le d}$, respectively.

For positive exchangeable correlation structures, the MVN rectangle probabilities  (\ref{MVNrectangle})
can be quickly
computed to a desired accuracy that is $10^{-6}$ or less, because
the $d$-dimensional integrals
conveniently reduce to 1-dimensional integrals \cite[p. 48]{Johnson&Kotz72}. 
For general correlation structures,    there are several papers in the literature, e.g., \cite{joe95,genz92,genz&bretz02}, that focus on the
computation of the MVN rectangle probabilities,  and, conveniently,  the implementation of the proposed algorithms is available in contributed R packages \cite{Joe-Wei-11,genz-etal-2012}. For more details see  \cite{nikoloulopoulos13b}.

\subsection*{Illustrations for $d=4$ and technical details}
For $d=4$ the matrices involved in the calculation of the sensitivity matrix $\bfD_\g$ of the composite score functions $\g$ take the form:

$$
-\bfD_{\g_1}=\x_i^\top E\begin{pmatrix}
\frac{\p\s_{i1}^{(1)}(\bbf)}{\p\nu_{i1}}&0&0&0\\
0&\frac{\p\s_{i2}^{(1)}(\bbf)}{\p\nu_{i2}}&0&0\\
0&0&\frac{\p\s_{i3}^{(1)}(\bbf)}{\p\nu_{i3}}&0\\
0&0&0&\frac{\p\s_{i4}^{(1)}(\bbf)}{\p\nu_{i4}}
\end{pmatrix}\x_i;
$$

$$
-\bfD_{\g_{1,2}}=E
\begin{pmatrix}
\frac{\p\s_{i,12}^{(2)}(\bbf,\rho_{12})}{\p\nu_{i1}} & \frac{\p\s_{i,12}^{(2)}(\bbf,\rho_{12})}{\p\nu_{i2}} &0 &0\\
\frac{\p\s_{i,13}^{(2)}(\bbf,\rho_{13})}{\p\nu_{i1}} & 0& \frac{\p\s_{i,13}^{(2)}(\bbf,\rho_{13})}{\p\nu_{i3}} &0\\
\frac{\p\s_{i,14}^{(2)}(\bbf,\rho_{14})}{\p\nu_{i1}} & 0& 0&\frac{\p\s_{i,14}^{(2)}(\bbf,\rho_{14})}{\p\nu_{i4}}\\
0&\frac{\p\s_{i,23}^{(2)}(\bbf,\rho_{23})}{\p\nu_{i2}}&\frac{\p\s_{i,23}^{(2)}(\bbf,\rho_{23})}{\p\nu_{i3}}&0\\
0&\frac{\p\s_{i,24}^{(2)}(\bbf,\rho_{24})}{\p\nu_{i2}}&0&\frac{\p\s_{i,24}^{(2)}(\bbf,\rho_{24})}{\p\nu_{i4}}\\
0&0&\frac{\p\s_{i,34}^{(2)}(\bbf,\rho_{34})}{\p\nu_{i3}}&\frac{\p\s_{i,34}^{(2)}(\bbf,\rho_{34})}{\p\nu_{i4}}
\end{pmatrix}\x_i;
$$

$$
-\bfD_{\g_2}=E
\begin{pmatrix}
\frac{\p\s_{i,12}^{(2)}(\bbf,\rho_{12})}{\p\rho_{12}} & 0&0&0&0&0\\
0&\frac{\p\s_{i,13}^{(2)}(\bbf,\rho_{13})}{\p\rho_{13}} &0&0&0&0\\
0&0& \frac{\p\s_{i,23}^{(2)}(\bbf,\rho_{14})}{\p\rho_{14}}&0&0&0\\
0&0&0& \frac{\p\s_{i,23}^{(2)}(\bbf,\rho_{23})}{\p\rho_{23}}&0&0\\
0&0&0&0& \frac{\p\s_{i,24}^{(2)}(\bbf,\rho_{24})}{\p\rho_{24}}&0\\
0&0&0&0&0& \frac{\p\s_{i,34}^{(2)}(\bbf,\rho_{34})}{\p\rho_{34}}
\end{pmatrix}.
$$
The elements of these matrices are calculated as below:
$$
-E\Bigl(\frac{\p\s_{i,jk}^{(2)}(\bbf,\rho_{jk})}{\p\rho_{jk}}\Bigr)=
-E\Bigl(\frac{\p^2\ell_2(\nu_{ij},\nu_{ik},\rho_{jk};y_{ij},y_{ik})}{\p\rho_{jk}^2}\Bigr)
=E\Bigl(\bigr(\frac{\p\ell_2(\nu_{ij},\nu_{ik},\rho_{jk};y_{ij},y_{ik})}{\p\rho_{jk}}\bigl)^2\Bigr),
$$
where $\frac{\p\ell_2(\nu_{ij},\nu_{ik},\rho_{jk};y_{ij},y_{ik})}{\p\rho_{jk}}=
\frac{\p f_2(y_{ij},y_{ik};\nu_{ij},\nu_{ik},\rho_{jk})}{\p\rho_{jk}}/f_2(y_{ij},y_{ik};\nu_{ij},\nu_{ik},\rho_{jk}),$

\begin{eqnarray*}-E\Bigl(\frac{\p\s_{i,jk}^{(2)}(\bbf,\rho_{jk})}{\p\bbf}\Bigr)
&=&-E\Bigl(\frac{\p^2\ell_2(\nu_{ij},\nu_{ik},\rho_{jk};y_{ij},y_{ik})}{\p\bbf\p\rho_{jk}}\Bigr)\\
&=&E\Bigl(\frac{\p\ell_2(\nu_{ij},\nu_{ik},\rho_{jk};y_{ij},y_{ik})}{\p\bbf}\frac{\p\ell_2(\nu_{ij},\nu_{ik},
\rho_{jk};y_{ij},y_{ik})}{\p\rho_{jk}}\Bigr);
\end{eqnarray*}

$\frac{\p\ell_2(\nu_{ij},\nu_{ik},\rho_{jk};y_{ij},y_{ik})}{\p\bbf}=
\frac{\p\log f_2(y_{ij},y_{ik};\nu_{ij},\nu_{ik},\rho_{jk})}{\p\bbf}=
\frac{\p f_2(y_{ij},y_{ik};\nu_{ij},\nu_{ik},\rho_{jk})}{\p\bbf}/f_2(y_{ij},y_{ik};\nu_{ij},\nu_{ik},\rho_{jk})$,

$\frac{\p f_2(y_{ij},y_{ik};\nu_{ij},\nu_{ik},\rho_{jk})}{\p\bbf}=
\frac{\p f_2(y_{ij},y_{ik};\nu_{ij},\nu_{ik},\rho_{jk})}{\p\nu_{ij}}\x_{ij} + \frac{\p f_2(y_{ij},y_{ik};\nu_{ij},\nu_{ik},\rho_{jk})}{\p\nu_{ik}}\x_{ik}$,

$\frac{\p f_2(y_{ij},y_{ik};\nu_{ij},\nu_{ik},\rho_{jk})}{\p\nu_{ij}}=
\frac{\p f_2(y_{ij},y_{ik};\nu_{ij},\nu_{ik},\rho_{jk})}{\p \Phi^{-1}\bigl(F_1(y_{ij};\nu_{ij})\bigr)}\frac{\p\Phi^{-1}\bigl(F_1(y_{ij};\nu_{ij})\bigr)}{\p \nu_{ij}}+\frac{\p f_2(y_{ij},y_{ik};\nu_{ij},\nu_{ik},\rho_{jk})}{\p \Phi^{-1}\bigl(F_1(y_{ij}-1;\nu_{ij})\bigr)}\frac{\p\Phi^{-1}\bigl(F_1(y_{ij}-1;\nu_{ij})\bigr)}{\p \nu_{ij}}
$,

$\frac{\p\Phi^{-1}\bigl(F_1(y_{ij};\nu_{ij})\bigr)}{\p \nu_{ij}}=
\sum_{0}^{y_{ij}}\frac{\p f_1(y_{ij};\nu_{ij})}{\p \nu_{ij}}/{\phi\Bigl(\Phi^{-1}\bigl(F_1(y_{ij};\nu_{ij})\bigr)\Bigr)}$,
$\frac{\p f_1(y_{ij};\nu_{ij})}{\p \nu_{ij}}=f_1(y_{ij};\nu_{ij})\frac{\p\ell_1(\nu_{ij};y_{ij})}{\p\nu_{ij}}.$

\noindent The derivatives $\frac{\p f_2(y_{ij},y_{ik};\nu_{ij},\nu_{ik},\rho_{jk})}{\p\rho_{jk}}$ and $\frac{\p f_2(y_{ij},y_{ik};\nu_{ij},\nu_{ik},\rho_{jk})}{\p \Phi^{-1}\bigl(F_j(y_{ij};\nu_{ij})\bigr)}$ are computed with the  R functions {\it exchmvn.deriv.rho} and {\it exchmvn.deriv.margin}, respectively, in the  R package {\it mprobit} \cite{Joe-Wei-11}.

For $d=4$ the matrices involved in the calculation of the covariance matrix $\M_\g$ of the composite score functions $\g$ take the form:
\begin{footnotesize}
$$
\O_i^{(1)}=
\begin{pmatrix}
\mbox{Var}\Bigl(\s_{i1}^{(1)}(\bbf)\Bigr) & \mbox{Cov}\Bigl(\s_{i1}^{(1)}(\bbf),\s_{i2}^{(1)}(\bbf)\Bigr)&
\mbox{Cov}\Bigl(\s_{i1}^{(1)}(\bbf),\s_{i3}^{(1)}(\bbf)\Bigr)& \mbox{Cov}\Bigl(\s_{i1}^{(1)}(\bbf),\s_{i4}^{(1)}(\bbf)\Bigr)\\
\mbox{Cov}\Bigl(\s_{i2}^{(1)}(\bbf),\s_{i1}^{(1)}(\bbf)\Bigr)&\mbox{Var}\Bigl(\s_{i2}^{(1)}(\bbf)\Bigr)&
\mbox{Cov}\Bigl(\s_{i2}^{(1)}(\bbf),\s_{i3}^{(1)}(\bbf)\Bigr)&\mbox{Cov}\Bigl(\s_{i2}^{(1)}(\bbf),\s_{i4}^{(1)}(\bbf)\Bigr)\\
\mbox{Cov}\Bigl(\s_{i3}^{(1)}(\bbf),\s_{i1}^{(1)}(\bbf)\Bigr)&\mbox{Cov}\Bigl(\s_{i3}^{(1)}(\bbf),\s_{i2}^{(1)}(\bbf)\Bigr)&
\mbox{Var}\Bigl(\s_{i3}^{(1)}(\bbf)\Bigr)&\mbox{Cov}\Bigl(\s_{i3}^{(1)}(\bbf),\s_{i4}^{(1)}(\bbf)\Bigr)\\
\mbox{Cov}\Bigl(\s_{i4}^{(1)}(\bbf),\s_{i1}^{(1)}(\bbf)\Bigr)&\mbox{Cov}\Bigl(\s_{i4}^{(1)}(\bbf),\s_{i2}^{(1)}(\bbf)\Bigr)&
\mbox{Cov}\Bigl(\s_{i4}^{(1)}(\bbf),\s_{i3}^{(1)}(\bbf)\Bigr)&\mbox{Var}\Bigl(\s_{i4}^{(1)}(\bbf)\Bigr)
\end{pmatrix},
$$

$$
\O_i^{(1,2)}=\begin{pmatrix}
\mbox{Cov}\Bigl(\s_{i1}^{(1)},\s_{i,12}^{(2)}\Bigr) &
\mbox{Cov}\Bigl(\s_{i1}^{(1)},\s_{i,13}^{(2)}\Bigr) &
\mbox{Cov}\Bigl(\s_{i1}^{(1)},\s_{i,14}^{(2)}\Bigr) &
\mbox{Cov}\Bigl(\s_{i1}^{(1)},\s_{i,23}^{(2)}\Bigr) &
\mbox{Cov}\Bigl(\s_{i1}^{(1)},\s_{i,24}^{(2)}\Bigr) &
\mbox{Cov}\Bigl(\s_{i1}^{(1)},\s_{i,34}^{(2)}\Bigr)\\

\mbox{Cov}\Bigl(\s_{i2}^{(1)},\s_{i,12}^{(2)}\Bigr) &
\mbox{Cov}\Bigl(\s_{i2}^{(1)},\s_{i,13}^{(2)}\Bigr) &
\mbox{Cov}\Bigl(\s_{i2}^{(1)},\s_{i,14}^{(2)}\Bigr) &
\mbox{Cov}\Bigl(\s_{i2}^{(1)},\s_{i,23}^{(2)}\Bigr) &
\mbox{Cov}\Bigl(\s_{i2}^{(1)},\s_{i,24}^{(2)}\Bigr) &
\mbox{Cov}\Bigl(\s_{i2}^{(1)},\s_{i,34}^{(2)}\Bigr)\\

\mbox{Cov}\Bigl(\s_{i3}^{(1)},\s_{i,12}^{(2)}\Bigr) &
\mbox{Cov}\Bigl(\s_{i3}^{(1)},\s_{i,13}^{(2)}\Bigr) &
\mbox{Cov}\Bigl(\s_{i3}^{(1)},\s_{i,14}^{(2)}\Bigr) &
\mbox{Cov}\Bigl(\s_{i3}^{(1)},\s_{i,23}^{(2)}\Bigr) &
\mbox{Cov}\Bigl(\s_{i3}^{(1)},\s_{i,24}^{(2)}\Bigr) &
\mbox{Cov}\Bigl(\s_{i3}^{(1)},\s_{i,34}^{(2)}\Bigr)\\

\mbox{Cov}\Bigl(\s_{i4}^{(1)},\s_{i,12}^{(2)}\Bigr) &
\mbox{Cov}\Bigl(\s_{i4}^{(1)},\s_{i,13}^{(2)}\Bigr) &
\mbox{Cov}\Bigl(\s_{i4}^{(1)},\s_{i,14}^{(2)}\Bigr) &
\mbox{Cov}\Bigl(\s_{i4}^{(1)},\s_{i,23}^{(2)}\Bigr) &
\mbox{Cov}\Bigl(\s_{i4}^{(1)},\s_{i,24}^{(2)}\Bigr) &
\mbox{Cov}\Bigl(\s_{i4}^{(1)},\s_{i,34}^{(2)}\Bigr)\\

\end{pmatrix},
$$
\end{footnotesize}

where
\begin{eqnarray*}
\mbox{Cov}\Bigl(\s_{ij_1}^{(1)},\s_{i,j_1j_2}^{(2)}\Bigr)&=&\sum_{\y} \s_{ij_1}^{(1)} \s_{i,j_1j_2}^{(2)} f_2(y_{ij_1},y_{ij_2};\nu_{ij_1},\nu_{ij_2},\rho_{j_1j_2}),\\
\mbox{Cov}\Bigl(\s_{ij_1}^{(1)},\s_{i,j_2j_3}^{(2)}\Bigr)&=&\sum_{\y} \s_{ij_1}^{(1)} \s_{i,j_2j_3}^{(2)} f_3(y_{ij_1},y_{ij_2},y_{ij_3};\nu_{ij_1},\nu_{ij_2},\nu_{ij_3},\rho_{j_1j_2},\rho_{j_1j_3},\rho_{j_2j_3}),
\end{eqnarray*}

and
\begin{footnotesize}
$$
\O_i^{(2)}=\begin{pmatrix}
\mbox{Var}\Bigl(\s_{i,12}^{(2)}\Bigr) &
\mbox{Cov}\Bigl(\s_{i,12}^{(2)},\s_{i,13}^{(2)}\Bigr) &
\mbox{Cov}\Bigl(\s_{i,12}^{(2)},\s_{i,14}^{(2)}\Bigr) &
\mbox{Cov}\Bigl(\s_{i,12}^{(2)},\s_{i,23}^{(2)}\Bigr) &
\mbox{Cov}\Bigl(\s_{i,12}^{(2)},\s_{i,24}^{(2)}\Bigr) &
\mbox{Cov}\Bigl(\s_{i,12}^{(2)},\s_{i,34}^{(2)}\Bigr)\\

\mbox{Cov}\Bigl(\s_{i,13}^{(2)},\s_{i,12}^{(2)}\Bigr) &
\mbox{Var}\Bigl(\s_{i,13}^{(2)}\Bigr) &
\mbox{Cov}\Bigl(\s_{i,13}^{(2)},\s_{i,14}^{(2)}\Bigr) &
\mbox{Cov}\Bigl(\s_{i,13}^{(2)},\s_{i,23}^{(2)}\Bigr) &
\mbox{Cov}\Bigl(\s_{i,13}^{(2)},\s_{i,24}^{(2)}\Bigr) &
\mbox{Cov}\Bigl(\s_{i,13}^{(2)},\s_{i,34}^{(2)}\Bigr)\\

\mbox{Cov}\Bigl(\s_{i,14}^{(2)},\s_{i,12}^{(2)}\Bigr) &
\mbox{Cov}\Bigl(\s_{i,14}^{(2)},\s_{i,13}^{(2)}\Bigr) &
\mbox{Var}\Bigl(\s_{i,14}^{(2)}\Bigr) &
\mbox{Cov}\Bigl(\s_{i,14}^{(2)},\s_{i,23}^{(2)}\Bigr) &
\mbox{Cov}\Bigl(\s_{i,14}^{(2)},\s_{i,24}^{(2)}\Bigr) &
\mbox{Cov}\Bigl(\s_{i,14}^{(2)},\s_{i,34}^{(2)}\Bigr)\\

\mbox{Cov}\Bigl(\s_{i,23}^{(2)},\s_{i,12}^{(2)}\Bigr) &
\mbox{Cov}\Bigl(\s_{i,23}^{(2)},\s_{i,13}^{(2)}\Bigr) &
\mbox{Cov}\Bigl(\s_{i,23}^{(2)},\s_{i,14}^{(2)}\Bigr) &
\mbox{Var}\Bigl(\s_{i,23}^{(2)}\Bigr) &
\mbox{Cov}\Bigl(\s_{i,23}^{(2)},\s_{i,24}^{(2)}\Bigr) &
\mbox{Cov}\Bigl(\s_{i,23}^{(2)},\s_{i,34}^{(2)}\Bigr)\\

\mbox{Cov}\Bigl(\s_{i,24}^{(2)},\s_{i,12}^{(2)}\Bigr) &
\mbox{Cov}\Bigl(\s_{i,24}^{(2)},\s_{i,13}^{(2)}\Bigr) &
\mbox{Cov}\Bigl(\s_{i,24}^{(2)},\s_{i,14}^{(2)}\Bigr) &
\mbox{Cov}\Bigl(\s_{i,24}^{(2)},\s_{i,23}^{(2)}\Bigr) &
\mbox{Var}\Bigl(\s_{i,24}^{(2)}\Bigr) &
\mbox{Cov}\Bigl(\s_{i,24}^{(2)},\s_{i,34}^{(2)}\Bigr)\\

\mbox{Cov}\Bigl(\s_{i,34}^{(2)},\s_{i,12}^{(2)}\Bigr) &
\mbox{Cov}\Bigl(\s_{i,34}^{(2)},\s_{i,13}^{(2)}\Bigr) &
\mbox{Cov}\Bigl(\s_{i,34}^{(2)},\s_{i,14}^{(2)}\Bigr) &
\mbox{Cov}\Bigl(\s_{i,34}^{(2)},\s_{i,23}^{(2)}\Bigr) &
\mbox{Cov}\Bigl(\s_{i,34}^{(2)},\s_{i,24}^{(2)}\Bigr) &
\mbox{Var}\Bigl(\s_{i,34}^{(2)}\Bigr)\\
\end{pmatrix},
$$
\end{footnotesize}
where
\begin{eqnarray*}
\mbox{Var}\Bigl(\s_{i,j_1j_2}^{(2)}\Bigr)&=&\sum_{\y} \s_{i,j_1j_2}^{(2)}\s_{i,j_1j_2}^{(2)} f_2(y_{ij_1},y_{ij_2}),\\
\mbox{Cov}\Bigl(\s_{i,j_1j_2}^{(2)},\s_{i,j_1j_3}^{(2)}\Bigr)&=&\sum_{\y} \s_{i,j_1j_2}^{(2)} \s_{i,j_1j_3}^{(2)} f_3(y_{ij_1},y_{ij_2},y_{ij_3}),\\
\mbox{Cov}\Bigl(\s_{i,j_1j_2}^{(2)},\s_{i,j_3j_4}^{(2)}\Bigr)&=&\sum_{\y} \s_{i,j_1j_2}^{(2)} \s_{i,j_3j_4}^{(2)}f_4(y_{ij_1},y_{ij_2},y_{ij_3},y_{ij_4}); 
\end{eqnarray*}
the inner sum is taken over all possible vectors $\y$.

\subsection*{One dependence parameter}
Differentiating $L_2$ with respect to $\rho$ leads to the bivariate composite score function:
$$
\g_2=\frac{\p L_2}{\p \rho}= \sum_{i=1}^{n}\s_i^{(2)}(\bbf,\rho)= \mathbf{0},
$$
where $\s_i^{(2)}(\bbf,\rho)=\frac{\sum_{j<k} \p \ell_2(\nu_{ij},\nu_{ik},\rho;y_{ij},y_{ik})}{\p \rho}$ is scalar.

The sensitivity matrix $\bfD_\g$ of the composite score functions $\g$ is given as below,
$$
\bfD_\g=-E\Bigl(\frac{\p \g}{\p\thbf}\Bigr)=
\begin{pmatrix}
-E\Bigl(\frac{\p \g_1}{\p\bbf}\Bigr)&-E\Bigl(\frac{\p \g_1}{\p\rhbf}\Bigr)\\
-E\Bigl(\frac{\p \g_2}{\p\bbf}\Bigr)&-E\Bigl(\frac{\p \g_2}{\p\rhbf}\Bigr)
\end{pmatrix}=
\begin{pmatrix}
\bfD^{(1)}&\mathbf{0}\\
\bfD^{(2,1)}&D^{(2)}
\end{pmatrix},
$$
where $\bfD^{(1)}=\frac{1}{n}\sum_i^n\x_i^\top\Debf_i^{(1)}\x_i$ with
$\Debf_i^{(1)}=-E\Bigl(\frac{\p\s_{i}^{(1)}(\bbf)}{\p\nu_{i}}\Bigr)$,
$\bfD^{(2,1)}=\frac{1}{n}\sum_i^n\Debf_i^{(2,1)}\x_i$ with $\Debf_i^{(2,1)}=-E\Bigl(\frac{\p\s_{i}^{(2)}(\bbf,\rho)}{\p\bbf}\Bigr)$, and
$D^{(2)}=\frac{1}{n}\sum_i^n\Delta_i^{(2)}$ with $\Delta_i^{(2)}=-E\Bigl(\frac{\p\s_{i}^{(2)}(\bbf,\rho)}{\p\rho}\Bigr)$,
where,
\begin{eqnarray*}
-E\Bigl(\frac{\p\s_{i}^{(2)}(\bbf,\rho)}{\p\rho}\Bigr)&=&-E\Bigl({\p\frac{\sum_{j<k} \p \ell_2(\nu_{ij},\nu_{ik},\rho;y_{ij},y_{ik})}{\p \rho}}/{\p\rho}\Bigr)\\
&=&-E\Bigl({\sum_{j<k}\frac{ \p^2 \ell_2(\nu_{ij},\nu_{ik},\rho;y_{ij},y_{ik})}{\p \rho^2}}\Bigr)\\
&=&\sum_{j<k} -E\Bigl({\frac{ \p^2 \ell_2(\nu_{ij},\nu_{ik},\rho;y_{ij},y_{ik})}{\p \rho^2}}\Bigr)\\
&=&\sum_{j<k} E\Bigl(({\frac{ \p \ell_2(\nu_{ij},\nu_{ik},\rho;y_{ij},y_{ik})}{\p \rho}}
)^2\Bigr),\\
\end{eqnarray*}
and
\begin{eqnarray*}
-E\Bigl(\frac{\p\s_{i}^{(2)}(\bbf,\rho)}{\p\bbf}\Bigr)&=&-E\Bigl({\p\frac{\sum_{j<k} \p \ell_2(\nu_{ij},\nu_{ik},\rho;y_{ij},y_{ik})}{\p \rho}}/{\p\bbf}\Bigr)\\
&=&-E\Bigl({\sum_{j<k}\frac{ \p^2 \ell_2(\nu_{ij},\nu_{ik},\rho;y_{ij},y_{ik})}{\p \rho\p\bbf}}\Bigr)\\
&=&\sum_{j<k} -E\Bigl({\frac{ \p^2 \ell_2(\nu_{ij},\nu_{ik},\rho;y_{ij},y_{ik})}{\p \rho\p\bbf}}\Bigr)\\
&=&\sum_{j<k} E\Bigl({\frac{ \p \ell_2(\nu_{ij},\nu_{ik},\rho;y_{ij},y_{ik})}{\p \rho}}
{\frac{ \p \ell_2(\nu_{ij},\nu_{ik},\rho;y_{ij},y_{ik})}{\p \bbf}}\Bigr).\\
\end{eqnarray*}

The covariance matrix $\M_\g$ of the composite score functions $\g$ is given as below
$$\M_\g=\mbox{Cov}(\g)=
\begin{pmatrix}\mbox{Cov}(\g_1) & \mbox{Cov}(\g_1,\g_2)\\
\mbox{Cov}(\g_2,\g_1) & \mbox{Cov}(\g_2)
\end{pmatrix}=
\begin{pmatrix}\M_\g^{(1)}& \M_\g^{(1,2)}\\
\M_\g^{(2,1)}& M_\g^{(2)}\end{pmatrix}=
\sum_i\begin{pmatrix}\x_i^\top\O_i^{(1)}\x_i& \x_i^\top\O_i^{(1,2)}\\
\O_i^{(2,1)}\x_i& \Omega_i^{(2)}\end{pmatrix},
$$
where
$$\begin{pmatrix}\O_i^{(1)}& \O_i^{(1,2)}\\
\O_i^{(2,1)}& \Omega_i^{(2)}\end{pmatrix}=
\begin{pmatrix}\mbox{Cov}\Bigl(\s_i^{(1)}(\bbf)\Bigr) & \mbox{Cov}\Bigl(\s_i^{(1)}(\bbf),\s_i^{(2)}(\bbf,\rho)\Bigr)\\
\mbox{Cov}\Bigl(\s_i^{(2)}(\bbf,\rho),\s_i^{(1)}(\bbf)\Bigr) & \mbox{Cov}\Bigl(\s_i^{(2)}(\bbf,\rho)\Bigr)
\end{pmatrix};$$

\begin{eqnarray*}
\mbox{Cov}\Bigl(\s_i^{(1)}(\bbf),\s_i^{(2)}(\bbf,\rho)\Bigr)&=&\sum_{\y}\s_i^{(1)}(\bbf)\s_i^{(2)}(\bbf,\rho)f_d(\y_i),\\
\mbox{Cov}\Bigl(\s_i^{(2)}(\bbf,\rho),\s_i^{(2)}(\bbf,\rho)\Bigr)&=&\sum_{\y}\s_i^{(2)}(\bbf,\rho)\s_i^{(2)}(\bbf,\rho)f_d(\y_i).
\end{eqnarray*}

\end{document}